# A nonequilibrium statistical field theory of swarms and other spatially extended complex systems

Mark M. Millonas

Complex Systems Group, Theoretical Division and Center for Nonlinear Studies,
MS B258 Los Alamos National Laboratory, Los Alamos, NM 87545
& Santa Fe Institute, Santa Fe, NM

**Abstract**

A class of models with applications to swarm behavior as well as many other types of spatially extended complex biological and physical systems is studied. Internal fluctuations can play an active role in the organization of the phase structure of such systems. Consequently, it is not possible to fully understand the behavior of these systems without explicitly incorporating the fluctuations. In particular, for the class of models studied here the effect of internal fluctuations due to finite size is a renormalized *decrease* in the temperature near the point of spontaneous symmetry breaking. We briefly outline how these models can be applied to the behavior of an ant swarm.

In this paper I introduce a class of models which is in line with the basic processes acting in a variety of systems in nature, particularly biological ones. Some systems which fall into this class are insect swarms, swimming bacteria and algae,[6] physical trail formation, the evolution of river networks,[7] diffusive transport in polymeric materials,[1] population distribution models, various types of fractal growth phenomena,[13] and developmental morphogenesis.[11]

Here we study what will be called *stigmergic processes* as a generalization of the concept of stigmergy introduced by Grassé[3] in the context of collective nest building in social insects. The hypothesis of stigmergy, as described by Wilson[14], is that *it is the work already accomplished, rather than direct communication among nest mates, that induces the insects to perform additional labor.* The concept of stigmergy has also been invoked more recently in regards to swarm behavior.[12]

The more generalized idea of a stigmergic process is realized here in systems composed of three basic ingredients. The first ingredient is a *particle dynamics* which obeys a Markov process on some finite state space $\mathcal{X}$. The particle density $\rho(\mathbf{x}, \tau)$ obeys the Master equation

$$\frac{\partial \rho(\mathbf{x}, \tau)}{\partial \tau} = \int_{\mathcal{X}} \{W_\tau(\mathbf{x}|\mathbf{y})\rho(\mathbf{y}, \tau) - W_\tau(\mathbf{y}|\mathbf{x})\rho(\mathbf{x}, \tau)\} \, d^D\mathbf{y}, \qquad (1)$$

where $W_\tau(\mathbf{x}|\mathbf{y})$ is the probability density to go from state $\mathbf{y}$ to $\mathbf{x}$ at time $\tau$. The second element is a *morphogenetic field* $\sigma(\mathbf{x}, \tau)$, representing the environment which the particles both respond to, and act on. We will study one of the simplest situations, a fixed one-component pheromonal field which evolves according to

$$\frac{\partial \sigma(\mathbf{x}, \tau)}{\partial \tau} = -\kappa \, \sigma + \eta \, \rho, \qquad (2)$$

where $\kappa$ measures the rate of evaporation, breakdown or removal of the substance, and $\eta$ the rate of emission of the pheromone by the organisms. Lastly, some form of *coupling* is made between the particles and the field. This coupling takes the form of a behavioral function which describes how the particles move in response to the morphogenetic field, and in turn, how the particles act back on this field.

As we shall see, small changes in the microscopic behavior of the particles can result in large changes in the global behavior of the swarm, or particle field. This variability has significant implications not only for the behavioral response of the swarm to external stimuli, but also in the evolution of cooperative behavior. Wilson has remarked that an understanding of how this occurs would *constitute a technical breakthrough of exciting proportions, for it will then be possible, by artificially changing the probability matrices, to estimate the true amount of behavioral evolution required to go from [the behavior of] one species to ... that of another.*[14] He has further remarked that such large behavioral changes resulting from small changes in the individual dynamics would provide evidence that social behavior evolves at least as rapidly as morphology in social insects. This could provide an explanation why *behavioral diversity far outstrips morphological diversity at the level of species and higher taxonomic categories* in social insects.

In the region of a nonequilibrium phase transition the morphogenetic field, and hence the transition matrix, changes very slowly on scales typical of the particle field relaxation time since in this region the unstable modes will exhibit critical slowing down and will relax on a time scale much longer than the time scale of the stable modes. The particle modes are said to be *slaved* to the morphogenetic field, and can be adiabatically eliminated from the picture.[4] We obtain the stochastic order parameter equation

$$\frac{\partial \sigma(\mathbf{x}, \tau)}{\partial \tau} = \kappa\ \sigma + \eta \rho_s[\boldsymbol{\sigma}] + \eta\ g[\boldsymbol{\sigma}]\xi(\mathbf{x}, \tau), \tag{3}$$

where $\rho_s[\boldsymbol{\sigma}]$ is the quasi-stationary particle density, $g[\boldsymbol{\sigma}]$ is a function describing the fluctuations of the quasi-stationary particle density about its mean value, and $E\{\xi(\mathbf{x}, \tau)\} = 0$, $E\{\xi(\mathbf{x}, \tau)\xi(\mathbf{x}', \tau')\} = \delta(\mathbf{x} - \mathbf{x}')\delta(\tau - \tau')$. Since $\rho_s$ will depend on both the global state of the morphogenetic fields, and on the global boundary conditions, this is a globally coupled set of equations for the evolution of the morphogenetic fields. *Slaving of the particle field therefore allows an explicitly coupled global dynamics to emerge from the strictly local interactions of the model*, providing a key to how a globally integrated response may emerge from a system of locally acting agents.

The the fluctuations in the system are state dependent. In addition to amplifying an instability which exists in the absence of noise, this type of fluctuation can also produce transitions and ordered behavior in its own right. One of the consequence of this fact is that slaved particle field will constructively determine the self-organization properties of the systems *through its fluctuating properties*, as well as through quasi-stationary values. This is a fact which should be constantly be born in mind when studying such models.

For the purposes of this paper we will consider the case where the transition matrix takes the form $W(\mathbf{x}|\mathbf{y}) \propto f(\sigma(\mathbf{x})) p(|\mathbf{x} - \mathbf{y}|)$, where $f$ is some weighting function describing the effect of the field $\sigma$ on the motion of the particles, and $p(|\mathbf{x} - \mathbf{y}|)$ is a probability distribution of jumps of length $r = |\mathbf{x} - \mathbf{y}|$. Transition matrices of this type obey detailed balance, $W(\mathbf{x}|\mathbf{y})f(\sigma(\mathbf{y})) = W(\mathbf{y}|\mathbf{x})f(\sigma(\mathbf{x}))$. In this case we can define a partition function

$$Z = \left\{\frac{1}{V} \int d^D\mathbf{x}\ f(\sigma(\mathbf{x}))\right\}^N, \tag{4}$$

where $V$ is the volume of the state space $\mathcal{X}$, and $N$ is the total number of particles. A one-to-one analogy with a thermodynamic system with energy $U(\sigma(\mathbf{x}))$ and temperature $T = \beta^{-1}$ can be made if we set $f(\sigma(\mathbf{x})) = \exp(-\beta U(\sigma(\mathbf{x})))$, where any parameter $T$ can be regarded as a temperature parameter if $f(\sigma(\mathbf{x}); \alpha\ T) = f^{-\alpha}(\sigma(\mathbf{x}); T)$. Statistical quantities of interest can be calculated from the partition function according to the usual prescriptions. In a closed system the mean particle density and dispersion in the energy state $\epsilon$ are given by

$$E\{\rho_\epsilon\} = \frac{N}{VZ}\exp(-\beta\epsilon), \quad E\left\{(\Delta\rho_\epsilon)^2\right\} = \frac{E\{\rho_\epsilon\}}{\mu_\epsilon}\left(1 - \frac{\mu_\epsilon}{N}E\{\rho_\epsilon\}\right), \quad (5)$$

where $\mu_\epsilon$ is the volume of the system in energy state $\epsilon$. The slaved particle field in energy state $\epsilon$ can then be represented, to lowest order in the fluctuations, by $\rho_\epsilon[\boldsymbol{\sigma}] = E\{\rho_\epsilon[\boldsymbol{\sigma}]\} + \sqrt{E\{(\Delta\rho_\epsilon)^2[\boldsymbol{\sigma}]\}}\,\xi(\mathbf{x},t)$.

We introduce the dimensionless parameter $\bar{\rho} = N/V$, the mean density of particles, and $\upsilon = \mu^-/\mu^+$, the ratio of the volume of the field $\sigma(\mathbf{x})$ in the $\sigma^-$ state to the volume in the $\sigma^+$ state. We also define the function $R(\sigma^+, \sigma^-) = f(\sigma^+)/f(\sigma^-)$. In the mean field approximation a Langevin equation

$$\frac{dm}{dt} = -m + F(m) + \frac{1}{\sqrt{N}}Q(m)\,\xi(t) \quad (6)$$

for the order parameter $m$ can be derived,[10] where

$$F = \bar{\rho}(1+\upsilon)\frac{R^\beta - 1}{R^\beta + \upsilon}, \quad Q^2 = \frac{\bar{\rho}^2(1+\upsilon)^4}{\upsilon}\frac{R^\beta}{(R^\beta + \upsilon)^2}, \quad (7)$$

and where the $F$ and $Q$ are determined as functions of $m$ by

$$R(m) = R\left(\bar{\rho} + \frac{\upsilon\,m}{1+\upsilon}, \bar{\rho} - \frac{m}{1+\upsilon}\right). \quad (8)$$

The order parameter $m$ is analogous to a gas-liquid order parameter, and represents the difference in the values of the field in the $\sigma^+$ and $\sigma^-$ states after spontaneous symmetry breaking. The behavior of this system is described by the potential function

$$\Phi(m) = \int^m \frac{m - F(m)}{Q^2(m)}\,dx + \frac{1}{N}\ln Q, \quad (9)$$

where the phases $m_i$ of the system are determined by the conditions $\Phi'(m_i) = 0$, $\Phi''(m_i) > 0$.[5]

In the continuum limit $(N \to \infty)$ it can be shown that the critical value of the mean density $\bar{\rho}_c$ at which spontaneous symmetry breaking occurs is given by the condition $-\bar{\rho}_c\,U'(\bar{\rho}_c) = T$. Generally $\bar{\rho}_c$ is will increase with increasing temperature. The relative stability of two phases $m_1$ and $m_2$ is determined by the relative potentials $\Phi(m_1)$ and $\Phi(m_2)$ for each phase. Even in the continuum limit the details of the fluctuations cannot be neglected due to the presence of the factor $Q^2(m)$ under the integral in 9, and the relative stability of the phases will depend on the precise details of the internal fluctuations. Similar observations have been made elsewhere by Landauer and others.[8]

When $N$ is finite, the situation is still more complicated. It is clear that the possible values of the order parameter and the phase structure do not remain unchanged under the influence of internal fluctuations. The criterion for spontaneous symmetry breaking in this case is $-\bar{\rho}_c\, U'(\bar{\rho}_c) = \hat{T}$, where $\hat{T}$ is the renormalized temperature $\hat{T} = \gamma(N)T$ where $\gamma(N) = \sqrt{N + (N/2)^2} - N/2$. This is precisely the continuum condition except that the finite size fluctuations have the effect of renormalizing the temperature by the factor $\gamma(N) < 1$. The effect of increasing the internal fluctuations through decreasing the total number of particles has the effect of *decreasing the temperature*. We thus arrive at the seeming paradox that increased internal fluctuations may produce increased order.

I will now briefly outline how the previous analysis can be applied to the example of an ant swarm. More details can be found elsewhere.[9] In this case the individual ants are the particles, and the morphogenetic field is a pheromonal substance which the ants sense with their antennae, and emit from their bodies as they move. The basic measurement the ants make is the quantity of pheromone receive by each antennae. They can therefore respond to difference in the pheromone between the antennae, and move accordingly. A very general model of such motion assume that the particle experience a *force* which is proportional to the scent gradient at that point multiplied by some nonlinear response function $\chi(\sigma)$ of the scent at that point. The nonlinear response function models the nonlinearities underling the basic physiology of the sensing apparatus, for instance, any nonlinear neural/receptor response to the pheromone, including such effects as saturation of the receptor sites on the antennae by the pheromonal substance. In addition there is an element of randomness due to fluctuations in the external environment as well as internal fluctuations. These are incorporated into an effective random force with a strength in proportion to $\sqrt{T}$ where $T$ is the temperature factor. The motion of a particle can be described by a Langevin equation of the form

$$\frac{d\mathbf{x}(t)}{dt} = \chi(\sigma(\mathbf{x}))\nabla\sigma + \sqrt{2T}\,\xi(t), \qquad (10)$$

where $E\{\xi(t)\} = 0$, and $E\{\xi(t)\xi(t')\} = \delta(t-t')$. This can be written in the

form
$$\frac{d\mathbf{x}(t)}{dt} = -\nabla U(\mathbf{x}) + \sqrt{2T}\,\xi(t), \tag{11}$$

where $\chi(\sigma(\mathbf{x})) = -U'(\sigma(\mathbf{x}))$. Easy to show that the behavioral function of such a system is given by $f(\sigma) = \exp(-\beta U(\sigma))$.

The microscopic dynamics of the ants which we will study in the rest of the paper is determined by the response function

$$\chi(\sigma) = \alpha + \frac{c\rho}{c+\rho}, \tag{12}$$

where $\rho$ is the pheromone density, and $\alpha$ and $c$ are constants with the units of pheromone density. This function is inspired by the observed behavior of actual ants[2]. The constant $\alpha$ is roughly the threshold where the response of the ants to the pheromone is small unless $\rho > \alpha$. The constant $c$ will be known as the capacity. When $\rho$ approaches $c$ the ants respond less accurately to pheromone gradients. This is because when the pheromone density is very large the antennae receptors become saturated and the ant can not sense the pheromone gradient as accurately.

For simplicity we will introduce the dimensionless variable $\sigma = \rho/\alpha$ and the dimensionless parameter $\delta = \alpha/c$, where $1/\delta$ is the dimensionless capacity. The energy function takes the form

$$U(\sigma) = -\ln\left(1 + \frac{\sigma}{1+\delta\sigma}\right), \tag{13}$$

where we drop off any additive constant term, which have no effect on the behavior of the ants. For the case where the density of ants it low, and hence the pheromone density is low ($\rho << g$), we can make use of the approximate energy function $U_0(\sigma) = -\ln(\alpha + \sigma)$.

An illustration of this effect is shown in Figure 1. A given current of organisms $I$ flows into a junction from the left. On the lower branch the pheromone density is fixed at $\sigma_0$, and on the upper branch $\sigma$ is allowed to vary. $\mathcal{T}(\sigma)$, the proportion of the current which flows into the upper branch, is given by the sigmoidal function

$$\mathcal{T}(\sigma) = \left[1 + \exp\left(\beta U(\sigma)/U(\sigma_0)\right)\right]^{-1}. \tag{14}$$

The plots on the right of Figure 1 shows $\mathcal{T}(\sigma)$ for varying values of $\beta$ and $\delta$. The upper plot, where $\delta$ is fixed, shows the influence of increasing the temperature (lowering $\beta$). As the temperature increases the threshold response becomes less and less pronounced. In the opposite limit $\beta \to \infty$, $\mathcal{T}(\sigma)$ would be a step function $\Theta(\sigma - \sigma_0)$. In this limit all of the ants would choose the branch with the greatest pheromone density. In the lower plot the noise level is fixed, and the capacity $c = \alpha/\delta$ is varied. It is interesting to note that the effects of decreasing the capacity with fixed temperature are similar to the effects of increasing the temperature with fixed capacity. When the density of the ants increases, the pheromone density increases up to and beyond the capacity, the qualitative effects on the behavior of the ants is the same *as if the temperature was increased*. This gives the swarm roughly the ability to modulate its temperature by modulating its numbers.

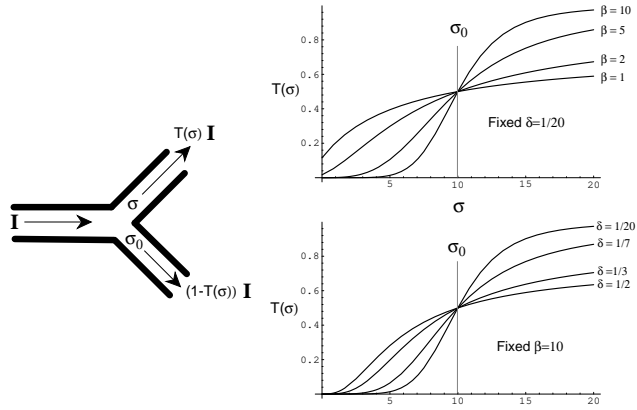

Figure 1: Transition functions for varying $\beta$ and $\delta$.

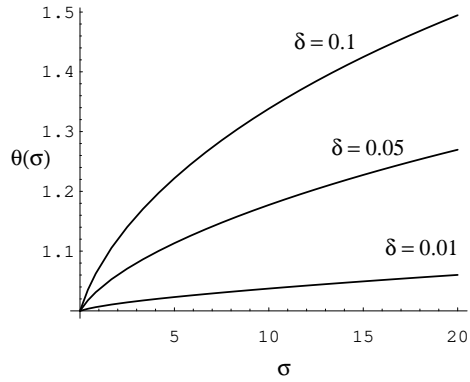

Figure 2: Effective temperature factor

This can be made more clear by defining an effective temperature factor $\theta(\sigma)$ through the relation $f(\sigma) = \exp(-\beta U_0(\sigma)/\theta(\sigma))$. $\theta(\sigma)$ roughly mea-sures the effective change in temperature as a function of the pheromonal field when compared to the case where $\delta = 0$, which correspond to the energy function $U_0$. The effective temperature is then given by $\theta(\sigma)T$ where

$$\theta(\sigma) = \frac{\ln\left(1 + \frac{\sigma}{1+\delta\sigma}\right)}{\ln(1+\sigma)}. \tag{15}$$

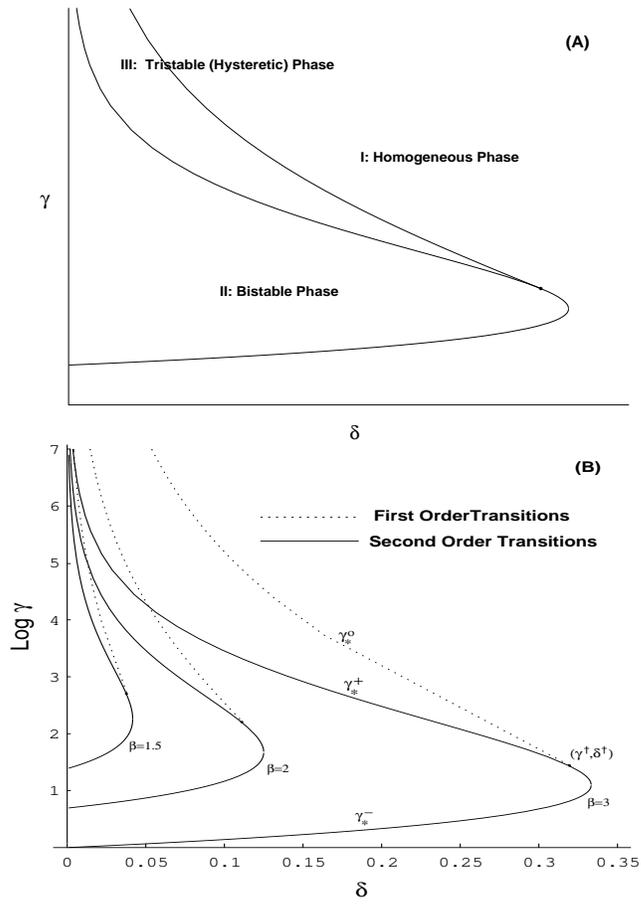

Figure 3: $\bar{\rho} - \delta$ phase diagram for the ant swarm.

Fig. 2 illustrates the increase in the effective temperature with increasing $\sigma$ for three different values of $\delta$. Since increasing the temperature tends to decrease stability, we might expect any organized behavior to breakdown when the number of participants grows too large. It is this ability of the swarm to self-modify its temperature which allows it, in a sense, to traverse its various phase transition boundaries.

Figure 3 is a typical phase plot for the ant swarm illustrating regions of homogeneity, bistability and hysteresis. The plot illustrates the effect of behavioral and swarm parameters on the swarm as a whole. In this case $\delta$ is a behavioral parameter which could be expected to change on the evolutionary time scale, and $\gamma$, which is proportional to the number of participants, is a swarm parameter which determines the behavioral "phase" of the swarm. More details may be found in previously published papers[9] where the properties of an ant swarm are analyzed in depth, and it is also shown how the collective behavior of real ants[2] can be understood in terms of such models.

# References

[1] R. W. Cox, & D. S. Cohen. *J. Polymer Sci. B* **27**, 589 (1989).

[2] J.-L. Deneubourg, S. Aron, S. Goss and J. M. Pasteels, *J. Insect Behavior* **3**, 159 (1990).

[3] P. P. Grassé, *Experientia* **15**, 356 (1959).

[4] H. Haken. *Synergetics*, Third Ed., Springer-Verlag (1983).

[5] W. Horsthemke, & R. Lefever, *Noise Induced Transitions*, Springer-Verlag (1984).

[6] J. O. Kessler. *Comments Theoretical Biology* **1**, 85 (1989).

[7] S. Kramer, & M. Marder. *Phys. Rev. Lett.* **68**, 205 (1992).

[8] R. Landauer, *J. Stat. Phys.* **53**, 233 (1988); N.G. Van Kampen, *IBM J. Res. Dev.* **32**, 107 (1988).

[9] M. M. Millonas, *J. Theor. Biol.* **159**, 529 (1992); In: *Cellular Automata and Cooperative Systems*, ( N. Bocara, E. Goles, S. Martinez & P. Pico, eds.), Kluwer (1993); In: *ALIFE III* (C.G. Langton, ed.), Santa Fe Institute: Addison-Wesley (1993).

[10] M. M. Millonas, Phys. Rev. E (to appear, 1993).

[11] J. E. Mittenthal. In: *Lectures in the Sciences of Complexity* (D. Stein, ed.), Addison-Wesley (1989).

[12] G. Theraulaz, & J.-L. Deneubourg. *SFI Working Paper* 92-09-046 (1992).

[13] T. Vicsek. *Fractal Growth Phenomena*, World Scientific (1989).

[14] E. O. Wilson, *The Insect Societies*, Belknap (1971).